\documentclass[12pt,preprint]{aastex}
\usepackage{psfig}

\shorttitle{The Giant Flare from SGR\,1806-20 and its Afterglow}
\shortauthors{Taylor \& Granot}
\begin{document}

\def\HI {H\kern0.1em{\sc i}}
\def\sgr{SGR~1806$-$20\ }
\def\simlt{\mathrel{\hbox{\rlap{\hbox{\lower4pt\hbox{$\sim$}}}\hbox{$<$}}}}
\def\simgt{\mathrel{\hbox{\rlap{\hbox{\lower4pt\hbox{$\sim$}}}\hbox{$>$}}}}

\title{The Giant Flare from SGR\,1806-20 and its Radio Afterglow}

\def\unm{1} \def\vlba{2}  \def\kipac{3}

\author{ G. B. Taylor \altaffilmark{\unm,\vlba} \& J. Granot \altaffilmark{\kipac}}

\altaffiltext{\unm}{University of New Mexico, Dept. of Physics and
Astronomy, Albuquerque, NM 87131, USA; gbtaylor@unm.edu}
\altaffiltext{\vlba}{National Radio Astronomy Observatory, Socorro, NM 87801, USA}
\altaffiltext{\kipac}{Kavli Institute of Particle Astrophysics and Cosmology, 
Stanford University, P.O Box 20450, MS 29, Stanford, CA, 94309, USA}

\begin{abstract}

 The multi-wavelength observations of the 2004 December~27 Giant Flare
(GF) from \sgr and its long-lived radio afterglow are briefly
reviewed. The GF appears to have been produced by a dramatic
reconfiguration of the magnetic field near the surface of the neutron
star, possibly accompanied by fractures in the crust.  The explosive
release of over $10^{46}\;$erg (isotropic equivalent) powered a
one-sided mildly relativistic outflow. The outflow produced a new
expanding radio nebula, that is still visible over a year after the
GF. Also considered are the constraints on the total energy in the GF,
the energy and mass in the outflow, and on the external density,
as well as possible implications for short $\gamma$-ray bursts and
potential signatures in high energy neutrinos, photons, or cosmic
rays. Some possible future observations of this and other GFs are
briefly discussed.

\end{abstract}

\keywords{pulsars: individual (SGR 1806-20) -- stars: neutron -- stars:flare
-- stars: winds,outflows -- radio continuum: general}

%\vfill\eject
\section{Introduction}

Magnetars are a small class of young, isolated, neutron stars with
extremely large magnetic fields of up to $\sim 10^{15}\;$G at the
surface \citep{dun92,pac92,kou98}. These objects give rise to
occasional bursts peaking in the hard X-ray to soft $\gamma$-ray
range, thus manifesting themselves as Soft Gamma-ray Repeaters (SGRs)
and Anomalous X-ray Pulsars (AXPs) \citep[for a review
see][]{WT04}. They have quiescent X-ray luminosities of $\sim
10^{34}-10^{35}\;{\rm erg\;s^{-1}}$, and are seen to pulse in the
X-rays with periods of $P \sim 5-12\;$s (the pulsed fraction of the
flux is $\sim 5-11\%$ for SGRs and $4-60\%$ for AXPs). The pulsation
period is identified with the rotational period of the star.  They
have rapid spin-down rates, $\dot{P} \sim 10^{-11}-10^{-10}\;{\rm
s\;s^{-1}}$, leading to age estimates of $P/2\dot{P} \sim
10^3-10^4\;$yr.  They have no detectable quiescent radio emission.
From the rapid spin-down, and lack of any signs of accretion (e.g.,
infrared excess, radio emission) {or orbital modulation of their
emission}, they are thought to be isolated (and not in binary or
multiple star systems).  Only 13 magnetars are known \citep{har05},
all in our Galaxy or the Large Magellenic Cloud (LMC) satellite
galaxy.
 
The SGRs are more boisterous compared to the AXPs, giving rise to
frequent bursts (with energies of $\lesssim 10^{41}\;$erg and typical
durations of $\sim 0.2\;$s), and on rare occasions, also emit giant
flares.  SGRs also have somewhat larger period derivatives, and
shorter spin-down ages, than AXPs.  SGR-like flares have been detected
from a few AXPs \citep{kas03}, supporting the notion that SGRs and
AXPs can be unified within the magnetar model.

Quite rarely, about once every $\sim 50\;$yr per source, SGRs 
produce a giant flare (GF). A GF consists of a very bright initial
spike that peaks in soft gamma rays (a few hundred keV) and lasts for
about a quarter of a second, followed by a longer and dimmer tail that
peaks in the hard X-rays ($\sim 10\;$keV) and lasts for a few hundred
seconds, with a strong modulation at the rotational period of the
neutron star.
So far only three GFs have been detected, originating from within our
galaxy or the LMC.  On 1979 March 5, SGR~0526$-$66
in the LMC produced the first GF that could be witnessed using
satellites in orbit about the Earth.  The peak (isotropic equivalent)
luminosity from this flare was estimated at $\sim 4\times
10^{44}\;{\rm erg\;s^{-1}}$, which exceeds the luminosity of the
entire Galaxy, for the 0.2 seconds of the peak emission. A fading tail
lasted 3 minutes and exhibited strong oscillations with a period of
8.1 seconds \citep{maz79}. The total isotropic equivalent energy
release was $\sim 5\times 10^{44}\;$erg. A similar GF event was
detected from SGR~1900$+$14
on 1998 August 27, with an isotropic equivalent peak luminosity and
total energy in excess of $3\times 10^{44}\;{\rm erg\;s^{-1}}$ and
$10^{44}\;$erg, respectively. Its initial spike and tail lasted for
$\sim 0.35\;$s and $\sim 400\;$s, respectively. A faint radio
afterglow was detected in the days following this GF \citep{fra99}. In
both of these two GFs the (isotropic equivalent) energy in the initial
spike was comparable to that in the tail.

There are also some intermediate events between the short, more
frequent, bursts and the giant flares, in terms of their total
duration and isotropic equivalent peak luminosity and energy
\citep[see][and references therein]{WT04}. They often occur following
giant flares. The largest so far of these intermediate events (on
2001 April 18, from SGR~1900+14) also showed several strong flux
modulations at the stellar rotational period. All this suggests that
there might be a continuum of events, differing mainly in their energy
release, and consequently also in their duration and peak
luminosity. It might make sense to roughly divide the different types
of such events according to their isotropic equivalent energy release,
$E_{\rm iso}$: (i) short, more frequent, bursts ($E_{\rm iso} \lesssim
10^{41}\;$erg), (ii) intermediate events ($10^{41}\;{\rm erg} \lesssim
E_{\rm iso} \lesssim 10^{44}\;$erg), and (iii) giant flares ($E_{\rm
iso} \gtrsim 10^{44}\;$erg).

The last and most spectacular of the three GFs we have witnessed so
far occurred on 2004 December 27 from GSR~1806$-$20.  The GF was
preceded by a gradual change in the spectral hardness, photon index,
and spin-down rate that peaked several months in advance of the GF
\citep{woods06}.  The sudden energy release of more than
$10^{46}\;$erg in gamma-rays (assuming isotropic emission at a
distance of $15\;$kpc) managed to eject a significant amount of
baryons, probably accompanied by some pairs and magnetic fields, from
the neutron star \citep{pal05,gel05,gra06}. As this outflow interacted
with the external medium, it powered an expanding radio nebula
\citep{cam05,gae05} at least 500 times more luminous than the only
other radio afterglow detected from an SGR GF \citep{fra99}.

We note that the energetics are reduced by a factor of 6$-$2 if one
adopts a lower distance of 6$-$10$\;$kpc found by \cite{cam05b} based
on \HI\ absorption observations.  However, a more detailed analysis by
\cite{MCC05} suggests that the \HI\ observations are consistent with
the X-ray absorption measurements that give a distance of 14.5 $\pm$
1.4 kpc \citep{cor97}, and with the distance of 15.1$^{+1.8}_{-1.3}$
kpc of the associated stellar cluster in G10.0$-$0.3 \citep{CE04}.
For this review we adopt a distance to \sgr of $15d_{15}\;$kpc, so
that $1\;$mas corresponds to $15d_{15}\;$AU or $2.25\times
10^{14}d_{15}\;$cm.

In this brief review we first consider how the GF was powered along with
implications for the neutron star
(\S~\ref{subsec:GF_obs}-\S~\ref{subsec:GF_QPOs}), for the possible
connection to short gamma-ray bursts (\S~\ref{subsec:GF_GRBs}), and
for possible signatures in high energy neutrinos, photons or cosmic
rays (\S~\ref{subsec:GF_HE}). We go on to discuss the radio
afterglow that is still being studied over a year after the GF
(\S~\ref{sec:radio}) and its implications for the properties of the
outflow from the GF and its environment.
We conclude with a short discussion of possible future work
(\S~\ref{sec:future}). 
The primary focus is on the observations and the immediate
insights gained from them.  A detailed discussion of the physics of
the GF is beyond the scope of this review.

\section{The Giant Flare Itself}

\subsection{Observational Highlights}
\label{subsec:GF_obs}

\begin{figure}
%%\epsscale{0.2}
%%\plotone{finf.eps}
%\special{psfile=f1.eps hoffset=-10 voffset=-50 hscale=80.0 vscale=80.0}
\vspace{-0.1cm}
\plotone{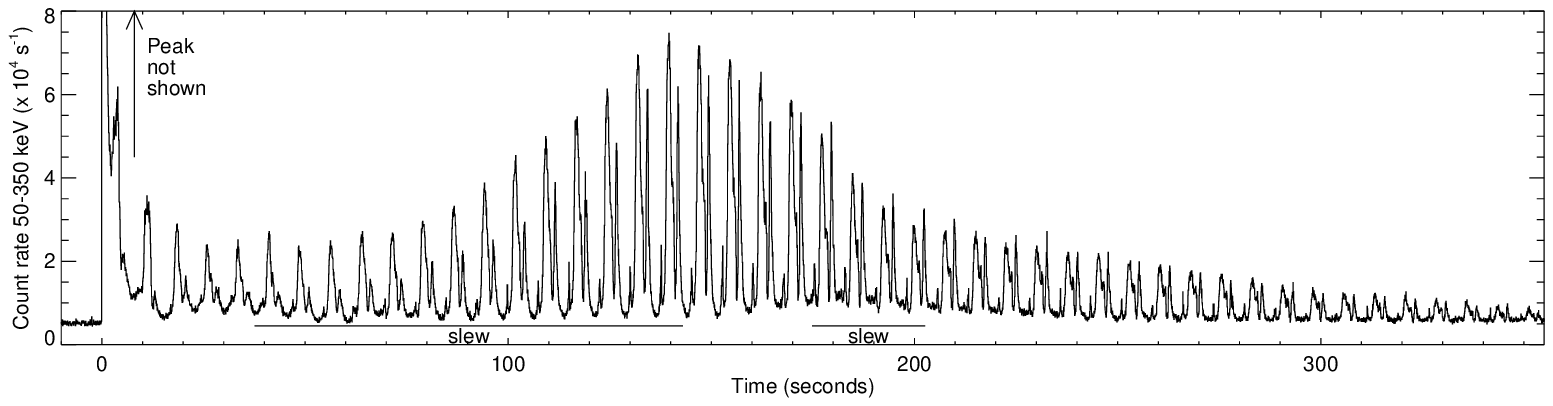}
\vspace{0.1cm}
\caption{Gamma-ray observations observed by SWIFT \citep[from][]{pal05} 
for the first 6 minutes after the GF from \sgr.
The peak of the emission is not shown, and some of the fluctuations in 
the count rate are due to the changing orientation of the spacecraft
as it was slewing.  }
\label{fig:gamma}
\end{figure}

The initial spike of the GF was the brightest astrophysical transient
event ever recorded (see Fig.~\ref{fig:gamma}, where its peak
saturated the instrument and is way-off the scale), surpassing even
the most intense solar flares \citep{tera05}. In particular, it was
the brightest blast of $\gamma$-rays detected in the $\sim 40$ years
that we have had detectors in orbit around the Earth. The total
fluence of $\sim 1-2\;{\rm erg\;cm^{-2}}$ saturated all but the least
sensitive particle detectors regardless of where the $\gamma$-ray
telescopes were pointed, and it created a disturbance in the Earth's
ionosphere \citep{camp05}. The (isotropic equivalent) luminosity of
the initial spike was $\sim (2-5) \times 10^{47}d_{15}^2\;{\rm erg\;s^{-1}}$
\citep{pal05,hur05,tera05}, or $\sim 10^3$ times that of the Galaxy,
and $\sim 500$ times more luminous than the two previous GFs. In fact,
it was so bright that even its echo off of the moon (due to Compton
scattering at the moon's surface) was detected \citep{maz05}. The
isotropic equivalent energy in the initial spike was $\sim (2-5)\times
10^{46}d_{15}^2\;$erg \citep{pal05,hur05,tera05} which is $\sim 100$
times larger than in the two previous GFs. The spectrum of the initial
spike was quasi-thermal (showing an exponential cutoff at high
energies and $F_\nu \propto \nu^{0.8}$ at low energies) with a
temperature of $kT \approx 0.2-0.5\;$MeV \citep{pal05,hur05}. A fading
tail was detected
with a strong modulation at the known rotational period of the neutron
star ($P = 7.56\;$s), with an isotropic equivalent energy release
[$\sim (0.5-1.2)\times 10^{44}d_{15}\;$erg] and duration ($\sim 380\;$s)
similar to the tails of the previous two GFs \citep{pal05,hur05}.

A precursor to the GF was detected $142\;$s before the initial spike,
and lasted for $\sim 1\;$s with a flat peak \citep{pal05,hur05}. The
initial spike was preceded by a gradual rise in the count rate which
reached a moderate value (of $\sim 3\times 10^4\;{\rm counts\;s^{-1}}$
with BAT on board {\it Swift}) within $40\;$ms, at which point the
main rise to the initial spike started with an exponential rise in the
flux by a factor of $\sim 10^3$ within $\sim 2\;$ms, corresponding to
an e-folding time of $0.3\;$ms. Thus the rise time was resolved for
the first time by {\it Swift} \citep{pal05}. After the initial
exponential rise there was at least one dip before the flux continued
to rise. Later, there is evidence for two more stages of exponential
flux increase with slower e-folding times of $5\;$ms and $70\;$ms
\citep{sch05}.

\subsection{The Basic Current Theoretical Framework}
\label{subsec:GF_theory}

The GF is believed to have originated from a violent magnetic
reconnection event in this magnetar \citep{TD95,TD01,pal05,hur05}.
The intense internal magnetic field of $\sim 10^{16}\;$G gradually
unwinds and transfers helicity across the stellar surface into the
magnetosphere.  The stresses on the crust gradually build up until a
fracture occurs, and the movement of the crust triggers a catastrophic
rearrangement of the external magnetic field. The rise time and the
duration of the initial spike are of the order of the Alfv\'en
crossing time of the magnetosphere and of the star, respectively,
while the duration of the tail corresponds to the cooling time of the
trapped fireball \citep{TD95,TD01,pal05,hur05}. The intermediate $\sim
5\;$ms time scale might be attributed to the propagation time of a
$\sim 5\;$km triggering fracture in the crust of the neutron star
\citep{sch05}. The observed temperature and isotropic equivalent
luminosity of the initial spike suggest that the energy was released
within a stellar radius or so from the surface of the neutron star
\citep{hur05,NPS05}.

The energy in the tail of the GF is believed to reflect the storage
capacity of the magnetosphere, and its near constancy between the
three GFs reflects the approximately similar magnetic energy between
these three magnetars. This is supported by the good fit to the
trapped fireball model \citep{TD95,TD01} of the time averaged (which
takes out the strong modulation at the rotational period of the
neutron star) light curve and spectrum of the tail for the GFs from
SGR~1900+14 \citep{Feroci01} and \sgr \citep{hur05}. Any excess energy
release during the initial spike, beyond the storage capacity of the
magnetosphere, is channeled either into radiation (mainly soft
$\gamma$-rays and hard X-rays), or into an outflow. Therefore, for
very energetic events most of the energy output comes out during the
initial spike, and thus the energy radiated in the initial spike can
vary dramatically between different GFs (not only for different
sources, as existing observations show, but also between different GFs
in the same source -- which cannot yet be tested observationally),
reflecting the diversity in the total energy release between different
events \citep{pal05,hur05}.

\subsection{Constraints on the Spin Down During the Giant Flare}
\label{subsec:GF_spindown}

Interestingly enough, there is no change in the spin or spin-down rate
associated with the GF \citep{pal05,woods05,woods06}. This is
consistent with the idea that the GF is powered by a reconfiguration
of the magnetic field and not by tapping the rotational energy (which
is insufficient anyway). Furthermore, this limits the amount of
angular momentum that could have been carried away by the outflow that
was launched during the GF. The extrapolation of the measurements
before and after the GF to the time of the GF gives $|\Delta\nu/\nu| <
5\times 10^{-6}$, which is significantly smaller than the spin-down of
$\Delta\nu/\nu \approx -1 \times 10^{-4}$ that was measured across the
1998 August 27 GF from SGR~1900+14 \citep{woods99}. The spin-down
during the initial spike of the 2004 December 27 GF from \sgr might
still be smaller than during its tail, despite the much larger energy
in the initial spike, because the much larger luminosity in the
initial spike reduces the radius out to which the outflowing material
co-rotates with the star, thereby reducing its specific angular
momentum \citep{Thompson00}. Nevertheless, the strict upper limit on
$|\Delta\nu/\nu|$ favors a relatively small outflowing mass, not much
larger than the lower limit (of $\gtrsim 10^{24.5}\;$g) that is
implied by the late time radio observations \citep{gel05,gra06}.

\subsection{QPOs in the Tail and Possible Fracture in the Crust}
\label{subsec:GF_QPOs}

Quasi-periodic oscillations (QPOs) with frequencies of $18$, $30.4$,
and $92.5\;$Hz have been detected in the oscillating tail of the GF
\citep{isr05} by RXTE, during part of the tail and over a certain
rotational phase of the neutron star. The $92.5\;$Hz QPO occurred
between $170$ and $220\;$s after the initial spike, in association
with a bump in the unpulsed component (corresponding to a reduction in
the amplitude of the pulsations at the rotational period of the
neutron star). The QPOs at $18$ and $92.5\;$Hz have been confirmed by
RHESSI \citep{WS06} which also found a stronger QPO at higher energies
with a frequency of $626.5\;$Hz that is visible at a different
rotational phase. Broadly similar QPOs were also found by RXTE in the
tail of the 1998 August 27 GF from SGR~1900+14 \citep{SW05}. Such QPOs
might arise from seismic modes in the neutrons star crust that drive
sheared Alfv\'en waves in the magnetosphere, and in particular
toroidal torsional modes that might be excited by a large scale
fracture of the crust \citep{isr05,SW05,Piro05,GSA06}, which had been
predicted to be excited in GFs (Duncan 1998; for a
different view see Levin 2006\nocite{Duncan98,Levin06}). The different rotational phase of the
$626.5\;$Hz and $92.5\;$Hz QPOs might suggest an origin in different
crustal fractures or magnetic reconnection events, the former
associated with the main flare and the latter with the late time
increase in the unpulsed emission \citep{WS06}.

\subsection{Possible Connection to Short-Hard Gamma-Ray Bursts}
\label{subsec:GF_GRBs}

The GF from \sgr could have been detected by BATSE out to a distance
of about $40\pm 10\;$Mpc \citep{pal05,hur05,nakar05}. At such large
extragalactic distances only the bright initial spike would be
detected, while the much dimmer pulsating tail would be below
detection threshold. Given that the initial spike has a duration,
variability, and energy spectrum roughly similar to gamma-ray bursts
(GRBs) of the short-hard class, this raises the possibility that some
fraction of the short GRBs are in fact extragalactic SGR GFs in
disguise \citep{hur05, pal05, nakar05, LGG05}. There are at least three
different lines of evidence which argue that the fraction $f_{\rm GF}$
of short-hard GRBs in the BATSE catalog that might be GFs from
extragalactic SGRs is small.

First, the lack of sufficiently bright host galaxies in the error
boxes of the six best localized short BATSE GRBs implies that these
events are fairly distant and more energetic than GFs (with an
isotropic equivalent energy release of $\gtrsim 10^{49}\;$erg), and
therefore $f_{\rm GF} \lesssim 0.15$ \citep{nakar05}. Second, if
indeed the birth rate of magnetars follows the star formation rate
(SFR; as is suggested by their relatively small inferred ages of a few
thousand years) we would expect to see an excess of events from the
direction of the Virgo galaxy cluster \citep{pal05} or of nearby star
forming galaxies \citep{PS06}. The lack of such an excess (there is no
apparent deviation from an isotropic distribution on the sky) implies
$f_{\rm GF} \lesssim 0.05$, and that either the Galactic rate of GFs
as luminous as the 2004 December 27 GF from \sgr is smaller (no more
than three per millennium) than might be expected naively from the
single such event that was detected so far (about three per century),
or that the distance to \sgr is $\sim 6-7\;$kpc instead of $15\;$kpc
\citep[which does not appear very likely;][]{MCC05}. A third line of
argument is based on the quasi-thermal spectrum of the initial spikes
of GFs, which has an exponential cutoff at high energies, in contrast
to the power law spectrum at high energies of almost all short-hard
BATSE GRBs \citep{LGG05}, which suggests $f_{\rm GF} \lesssim 0.04$.

Interestingly enough, there is some evidence for a correlation between
the directions of short-hard BATSE GRBs, and those of nearby galaxies
\citep{tan05}. Within $40\;$Mpc, this implies about $f_{\rm GF}
\approx 0.09^{+0.04}_{-0.03}$, which is marginally consistent with the
upper limits mentioned above. However, there is a stronger correlation
with early-type galaxies (which have a low current SFR) compared to
galaxies of all types, which is strange if this correlation is indeed
due to extragalactic SGR GFs.
This correlation, if true, might arise more naturally in models that
involve a long time delay between the star formation epoch and the
onset of short GRBs, such as binary mergers, if their luminosity
function is broad enough to account for the relatively low
luminosities of the required nearby events.

\subsection{High Energy Neutrinos, Photons, and Cosmic-rays}
\label{subsec:GF_HE}

The huge fluence of the 2004 December 27 GF from SGR~1806$-$20, of
$\sim 1-2\;{\rm erg\;cm^{-2}}$, is $\sim 10^4$ times larger than that
of the brightest recorded GRBs. This makes it an excellent candidate
for the detection of high-energy neutrinos
\citep{gel05,ioka05,hal05,eic05} and potentially also of ultra-high
energy cosmic rays \citep[UHECRs;][]{asano06} or high-energy photons
\citep{FZW05}.  High-energy neutrinos are expected to be produced in
internal shocks within the outflow, that arise due to variations in
its velocity, similar to the mechanism that had been proposed for GRBs
\citep{PX94,WB97}.  These mildly relativistic internal shocks are
believed to accelerate protons to high energies, which in turn produce
pions through $p$-$\gamma$ or $p$-$p$ interactions. As these pions
decay they produce high-energy neutrinos and photons. Some of the
shock accelerated protons may escape as UHECRs. In order for this
mechanism to work efficiently in SGR GFs, the outflow must be variable
and contain a significant amount of protons. Indeed, a significant
amount of protons in the outflow is implied by its large mass, that is
required in order to reproduce the extended coasting phase at a mildly
relativistic velocity (see \S \ref{subsec:dyn}), while a variable
outflow is suggested by the significant millisecond timescale
variability seen in the initial spike of the GF from SGR~1806$-$20
\citep{pal05}.

The expected neutrino event rates obviously depend on the model
assumptions. Most works have assumed a highly relativistic outflow,
while the radio observations suggest that at most $\sim 1\%$ of the
total energy was in such a highly relativistic component, and most of
the energy was in a mildly relativistic outflow
\citep{gra06}. Therefore, the expected event rates may require some
revisions. Recently, the IceCube collaboration has put out limits on
the flux of high-energy neutrinos and photons during the GF from
SGR1806$-$20, using the AMANDA-II detector \citep{Achtenberg06}.
These limits may constrain the conditions in the outflow from the GF
and call for further work. An interesting related prediction
\citep{ste05} is that if the internal toroidal magnetic field in newly
born rapidly rotating magnetars is large enough ($\gtrsim
10^{16.5}\;$G) then it would deform the star sufficiently such that
its gravitational wave signal might be detected by Advanced LIGO from
a magnetar as far away as the Virgo cluster. Such a high internal
magnetic field, however, is more than an order of magnitude larger
than that required in order to power giant flares \citep{Ofek06}.

\section{The Radio Afterglow of the Giant Flare}
\label{sec:radio}

\subsection{A One-Sided Mildly Relativistic Outflow}\label{subsec:radio}

\begin{figure}
%\vspace{20.0cm}
%%\epsscale{0.2}
%%\plotone{sizefit.eps}
%%\special{psfile=f2.eps angle=20 hoffset=-10 voffset=-50 hscale=50.0 vscale=50.0}
\vspace{-0.5cm}
\plotone{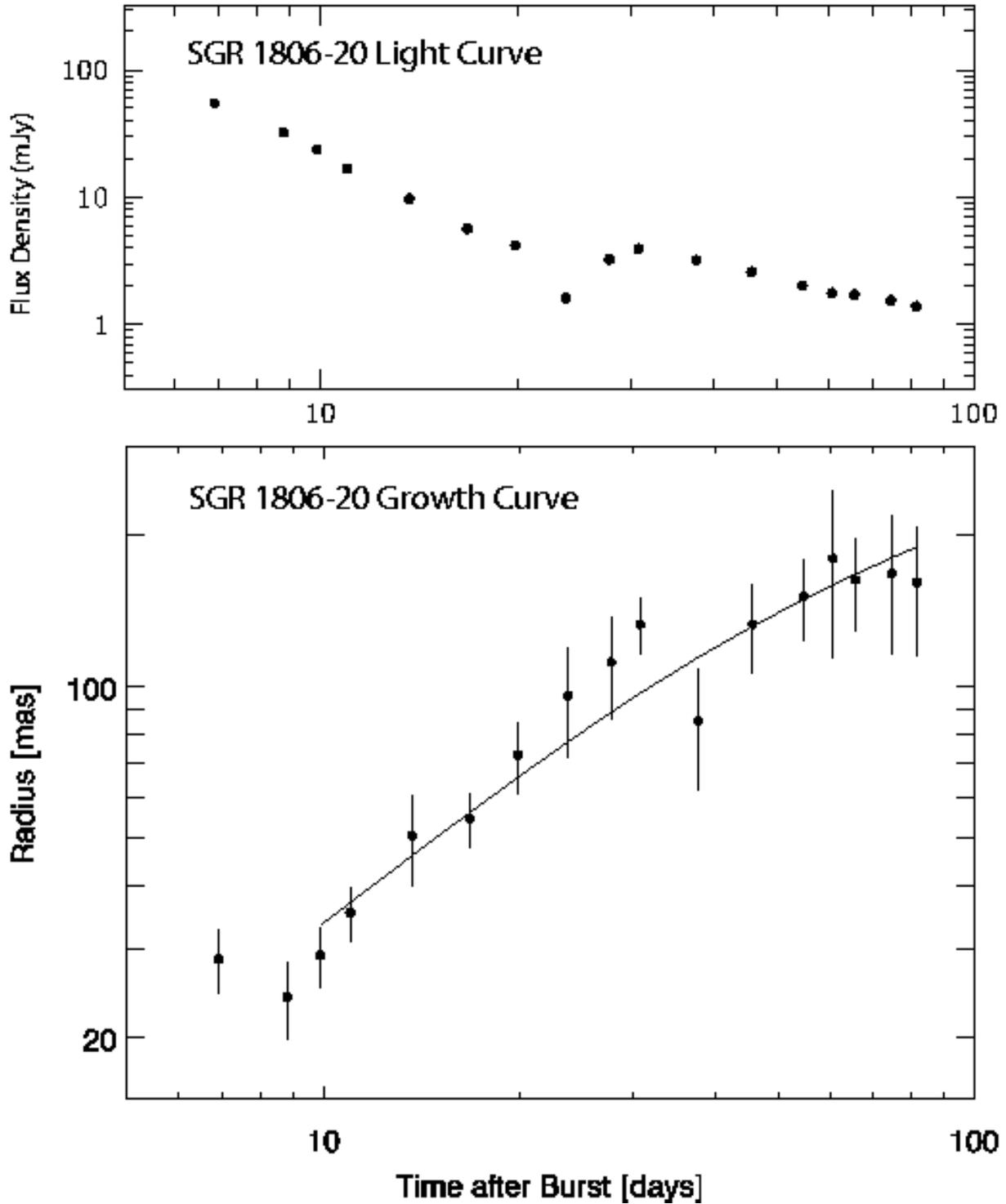}
\vspace{-1.0cm}
\caption{Expansion of the radio afterglow from \sgr as a function of
  time \citep[adapted from][]{tay05} (bottom panel).  The size shown is the geometric mean of
  the semi-major and semi-minor axes of the best fitting elliptical
  Gaussian for each observation. The solid line is a fit of a
  supersonically expanding shell model as described by Eq.~4 of
  \cite{gel05}. The top panel shows the 8.5 GHz light curve also
  from Taylor et al. 2005.}
\label{fig:sizemod}
\end{figure}

Radio observations of \sgr began $6.9\;$days after the GF, using the
Very Large Array (VLA) which fortuitously was in its largest, A
configuration, at the time.  The high angular resolution (0.2 arcsec)
afforded by the VLA in this configuration, together with its high
brightness, allowed the radio afterglow to be marginally resolved by
modelfitting to the visibility data.  The geometric mean size measured
was 57 mas, 7 days after the burst \citep{cam05b,gae05,tay05}.  After
30 days \citep[the time of a rebrightening reported by][]{gel05} the
radio afterglow had grown to $\sim260\;$mas.  Between 7 and 30 days
the growth of the radio nebula from 57 mas to 260 mas corresponds to
an average expansion velocity of 9.0 $\pm$ 1.6 mas/day (0.78 $\pm$
0.14 $d_{15}c$).  After this time, the growth rate appeared to slow
down (see Fig.~\ref{fig:sizemod}) so that the average expansion
velocity between day 30 and day 80 is 1.0 $\pm$ 2 mas/day ($<0.4
d_{15}c$) where the source size reached $\sim$322 mas \citep{tay05}.
As the radio afterglow was quite bright ($170\;$mJy at $1.4\;$GHz
after $7\;$days), it was also observed with a host of radio telescopes
including MERLIN and the Very Long Baseline Array (VLBA) to provide
even higher angular resolution \citep{fen06}.  These observations
revealed an elongated source with a $\sim 2:1$ axis ratio.  The
spectrum of the emission between 10 and 20 days after the GF is
well fit by a single power law with slope, $\alpha = -0.75 \pm 0.02$
(where $S_\nu \propto \nu^\alpha$) \cite{gae05}.  There is some 
evidence for a flatter spectrum before day 10 ($\alpha = -0.62 \pm 0.02$) and a steeper spectrum
after day 20 ($\alpha = -0.9 \pm 0.1$) \citep{cam05b}.  

\begin{figure}
%\vspace{20.0cm}
%\epsscale{0.7}
\vspace{-1.0cm}
\plotone{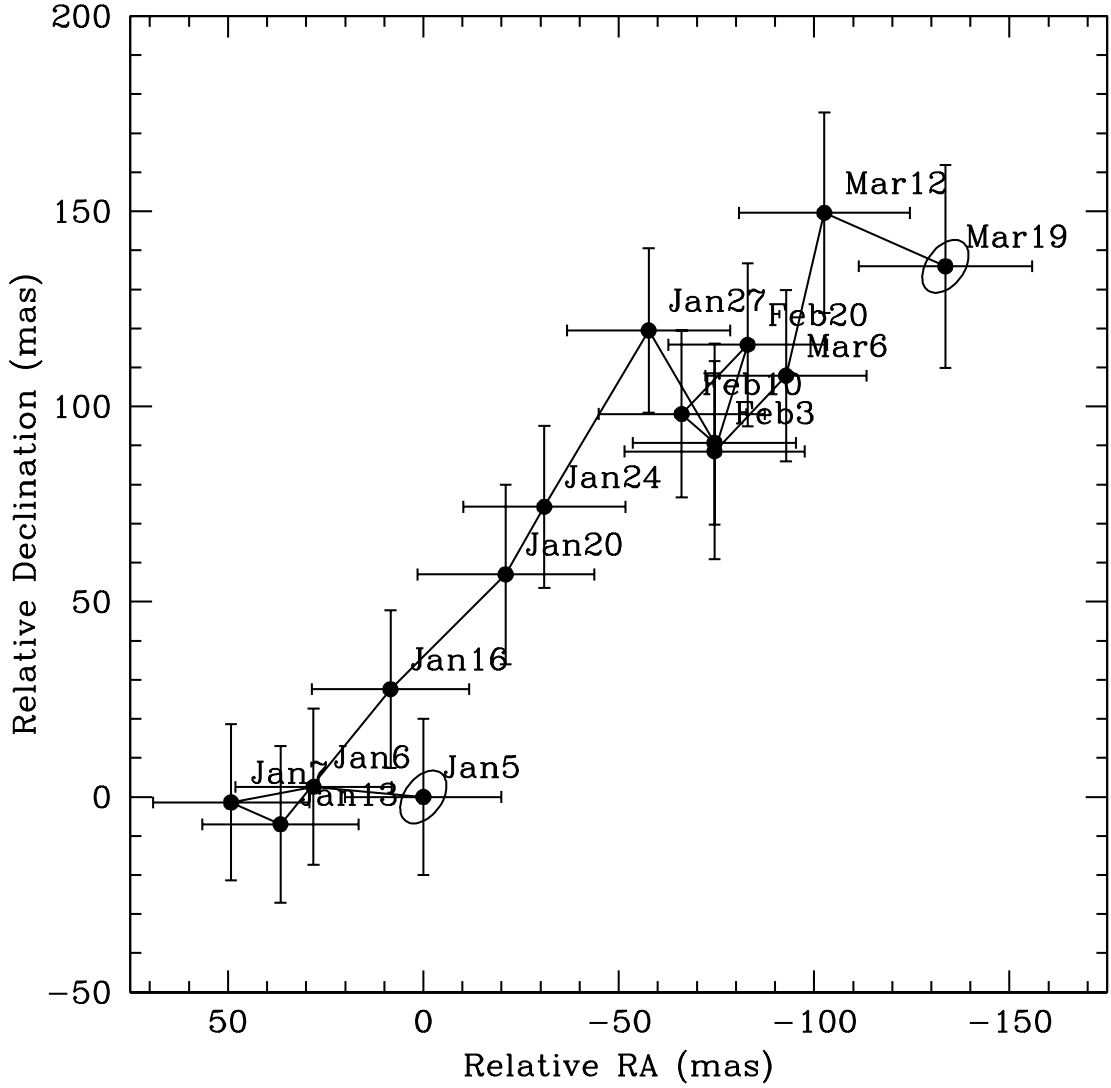}
\vspace{-1.8cm}
%\special{psfile=skyplot.eps hoffset=-10 voffset=-50 hscale=80.0 vscale=80.0}
\caption{The trajectory of the afterglow of \sgr
\citep[from][]{tay05}.  Dates are labeled.  The small ellipses denote
the first and last days used.}
\label{FIG3}
\end{figure}

\begin{figure}
%\epsscale{0.7}
\vspace{-1.0cm}
\plotone{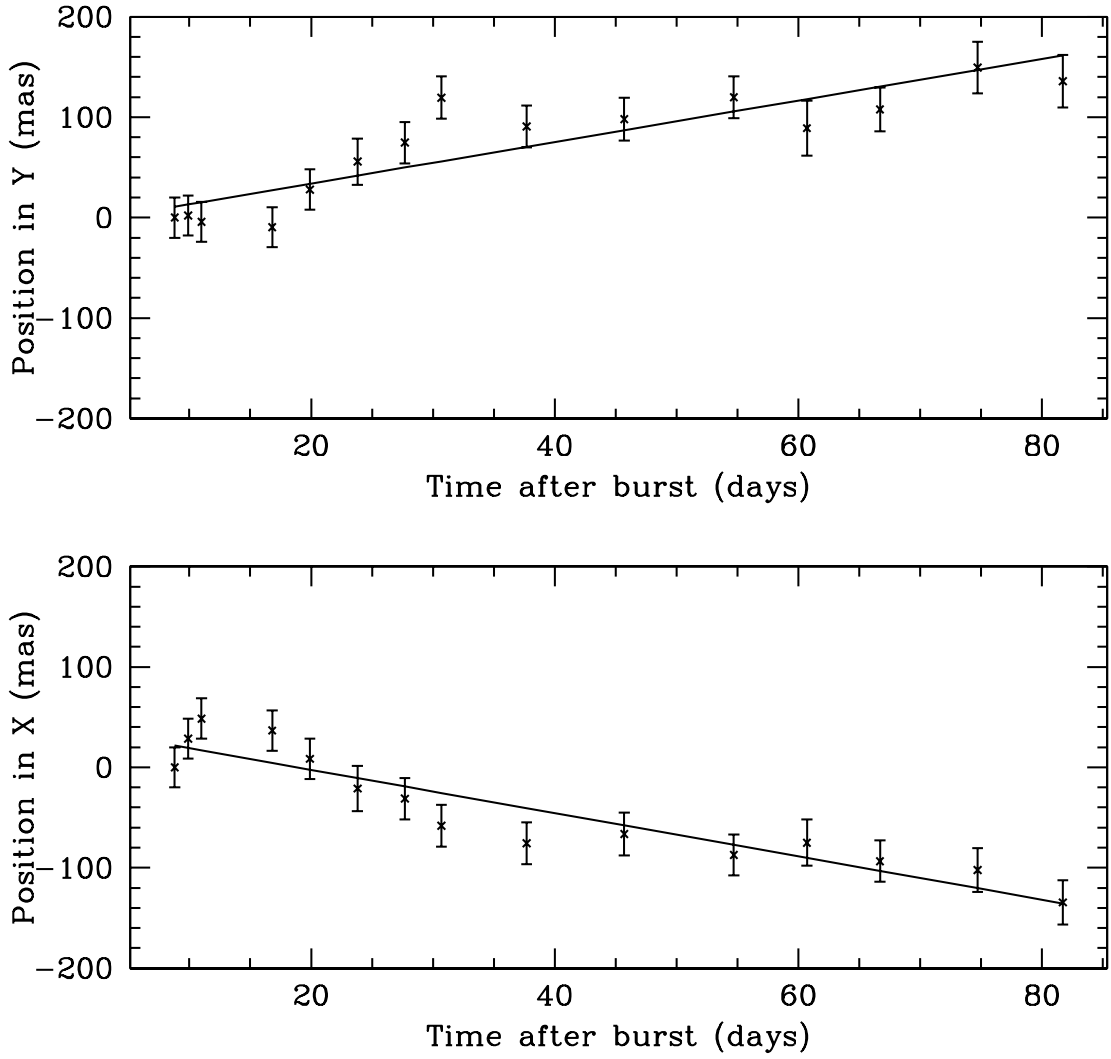}
\vspace{-1.8cm}
%\special{psfile=velx.eps hoffset=-10 voffset=-50 hscale=80.0 vscale=80.0}
\caption{Proper motion of the afterglow of \sgr.  The motion has been
decomposed into Right Ascension and Declination components of motion.  }
\label{FIG4}
\end{figure}

Furthermore, the centroid of the radio afterglow from \sgr was found
to shift by $\sim$200 mas over the course of the first 80 days
\citep[see Figs.~\ref{FIG3} and \ref{FIG4};][]{tay05}.  The radial
proper motion is 3.0 $\pm$ 0.34 mas/day at a position angle of $-$44
$\pm$ 6 $^\circ$ (measured north through east).  This motion
corresponds to $0.26 \pm 0.03 d_{15}c$.  There is some indication that
the time of fastest proper motion also corresponds to the time of
fastest growth.

\begin{figure}
\epsscale{0.8}
\vspace{14.0cm}
%\plotone{skytel.ps}
\vspace{-0.2cm}
\includegraphics{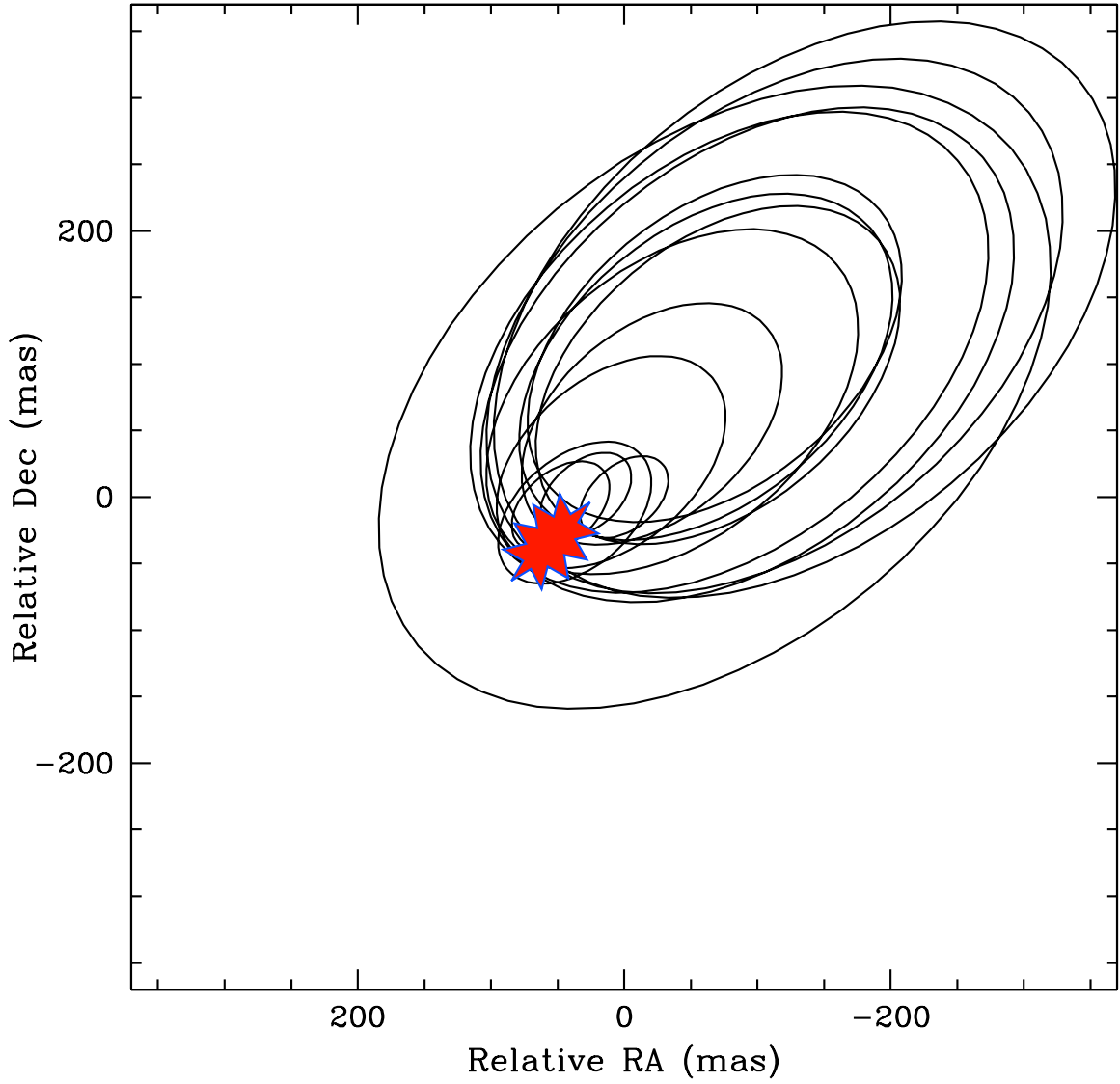}
\caption{A schematic of the growth and motion of the radio afterglow 
from \sgr with time.  The best fitting elliptical fits are drawn
for each epoch, properly centered on the derived position.  The position
of the initial explosion is illustrated in red.}
\label{FIG5}
\end{figure}

The motion of the radio flux centroid is along the major axis of the
source and is roughly half of the growth rate.  This may be naturally
explained by a predominantly one-sided outflow, which produces a radio
nebula extending from around the location of the magnetar out to a
particular preferred direction corresponding to the direction of the
ejection (Fig.~\ref{FIG5}).  This suggests that either the
catastrophic reconfiguration of the magnetic field which caused the GF
was relatively localized, rather than a global event involving the
whole magnetar \citep[c.f.,][]{eic02}, or that the baryonic content of
the ejecta is highly asymmetric.  The outflow must be intrinsically
one-sided since if there was a similar ``counter outflow'' in the
opposite direction, it should have produced significant radio
emission. The collision of the observed flow with the external shell
occurred around $t_{\rm col} \sim 5\;$days after the GF, so in order
not to detect emission from a counter outflow up to a time $t$ after
the GF, the distance of the shell (i.e. outer edge of the cavity) in
the opposite direction must be at least $t/t_{\rm col}$ times larger.
Therefore, the fact that there is no evidence for radio emission from
such a counter outflow up to hundreds of days ($\sim 100t_{\rm col}$)
after the GF would require an extreme asymmetry in the external
medium.

In the first 30 days, the leading edge of the one-sided expansion
moves away from the magnetar position at an apparent velocity of
$v_{\rm ap} \approx 0.8d_{15}c$ \citep{tay05}. The intrinsic velocity
is generally different and depends on the unknown inclination angle
$\theta$ of the outflow velocity at the apparent leading edge relative
to the line of sight. The minimum velocity is $v_{\rm min} \approx
0.62c$ for an inclination angle of $\theta_{\rm min} \approx
51^\circ$, and the true velocity is expected to be close to this value
\citep{gra06}. Interestingly enough, this is rather similar to the
escape velocity of $v_{\rm esc} \approx 0.5c$ from a neutron star.  At
these mildly relativistic velocities (minimal Lorentz factor
$\Gamma_{\rm min} \approx 1.3$) there is a modest increase in the
total kinetic energy for such a wide one-sided outflow compared to
simple estimates based on a spherical outflow \citep{gel05}. The total
kinetic energy increases by a factor of $\sim 2\,$--$\,3$, owing to
the factor $\sim 2$ higher velocity at the leading edge but lower
velocities elsewhere, while the isotropic equivalent kinetic energy
increases by a larger factor \citep{gra06}. This leads to a revised
estimate for the total kinetic energy in the ejecta of $\gtrsim
10^{44.5}\;$ergs. By momentum conservation, a one-sided outflow of
$10^{24.5}$ g \citep{gra06} at $0.62c$ imparts a kick to the magnetar
of $21(M_*/1.4\,M_\odot)^{-1}\;{\rm cm\;s^{-1}}$ where $M_*$ is the
mass of the neutron star (such a low kick velocity would be very hard
to detect).

The outflow does not remain (mildly) relativistic indefinitely.
Following \citet[][see their Eq. 4]{gel05}, the data from day 9
onwards is reasonably fit by a model featuring a supersonically
expanding spherical shell that is decelerated as it sweeps up material
\citep{tay05}.  While the deceleration of an anisotropic outflow might
be somewhat different than in the spherical case, the latter may still
serve as a rough approximation.  The fit (reduced $\chi^2$ of 0.76;
shown as the solid line in Fig.~\ref{fig:sizemod}) implies a
deceleration time of 40 $\pm$ 13 days after the GF, consistent with
the time of the peak rebrightening at $\sim 33\;$days (see upper panel
of Fig.~2).

\subsection{The Underlying Dynamical Model}
\label{subsec:dyn}

Here we present a simple dynamical model that can naturally account
for the radio observations \citep{gra06,gae05,gel05}. The reader is
referred to the literature for alternative views
\citep{wang05,yam05,dai05,lyu06}, which in our view are not as
successful in explaining all of the radio observations.

If the electrons that emit in the radio at the time of the first
observation ($6.9\;$days), at a distance of $\sim 10^{16}\;$cm from
the neutron star, had been accelerated near the neutron star (whose
radius is $\sim 10^6\;$cm), then they would have suffered huge
adiabatic losses, thus requiring an exceedingly large initial
energy. In addition, in the first 2$-$3 days of radio observations
(taken 7$-$9 days after the GF) the
flux was still rounding off ($\sim t^{-1.5}$) before reaching the
asymptotic steeper power law decay ($\sim t^{-2.7}$) that lasted until
$\sim 25\;$days.  This suggests that the radio emission lit up
slightly before the first observation (i.e, around $\sim 5\;$days), as
a result of a collision between the outflow that was ejected during
the initial spike of the GF and an external shell.  Such an external
shell naturally results due to the bow shock that is formed by the
quiescent relativistic pulsar-type wind of the neutron star as it
moves supersonically through the ambient medium \citep{gae05,gra06}.

\begin{figure}
%%\epsscale{0.2}
%%\plotone{finf.eps}
%\special{psfile=f1.eps hoffset=-10 voffset=-50 hscale=80.0 vscale=80.0}
%\vspace{-1.0cm}
\centerline{\includegraphics[width=10cm]{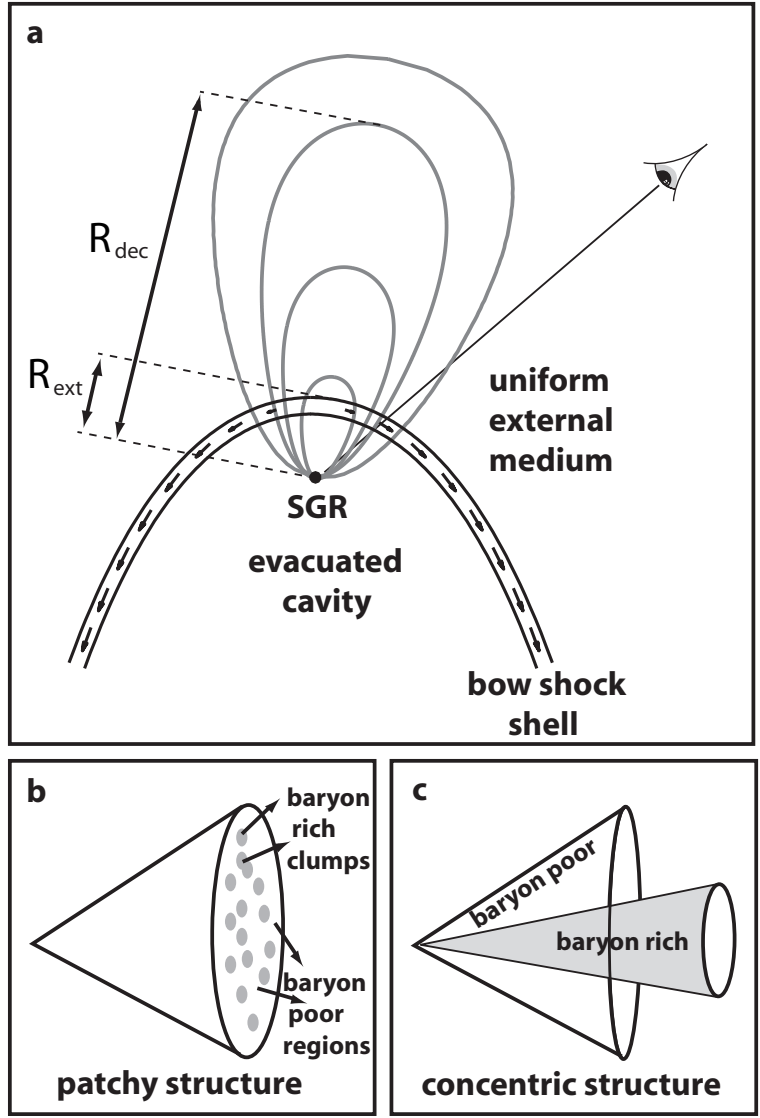}}
%\plotone{f6.eps}
%\vspace{-1.8cm}
\caption{Illustration of the basic underlying geometry in the
dynamical model \citep[from][]{gra06}. (a) A preexisting shell
surrounding a cavity (i.e., an evacuated region) is formed
due to the interaction of the SGR quiescent wind with the external
medium, and the SGR's supersonic motion relative to the external
medium. The outflow from the \sgr\ GF was
ejected mainly in one preferred direction, probably not aligned with
the head of the bow shock (which is in the direction of the SGR's
systemic motion). The ejecta collide with the external shell at a
radius $R_{\rm ext}$, and then the merged shell of shocked ejecta and
shocked swept up external shell continues to move outward at a constant
(mildly relativistic) velocity. As it coasts outward, it gradually
sweeps up the external medium until at a radius $R_{\rm dec} \sim
(4-5)R_{\rm ext}$ it has accumulated a sufficient mass 
to be significantly decelerated. At $R > R_{\rm dec}$ the structure
of the flow gradually approaches the spherical self-similar
Sedov-Taylor solution.  (b, c) Most of the mass in the outflow was in
baryons that were decoupled from the radiation, and our line
of sight was baryon-poor. This naturally occurs if there are separate
baryon-rich (radiation-poor) and baryon-poor (radiation-rich)
regions. Such regions might consist of small baryon-rich clumps
surrounded by baryon-poor regions (b) or might alternatively be part
of a global large-scale, possibly concentric configuration (c).
}
\label{fig:dyn_diag}
\end{figure}

During the collision the external shell is swept up by a forward shock
while the outflow is slightly decelerated by a reverse shock. After
the collision the merged shell keeps propagating outwards at a
constant coasting speed, and gradually sweeps up an increasing amount
of external medium. Initially the emission is dominated by the
electrons of the shocked shells. After the forward and reverse shocks
finish crossing these shells there is no fresh supply of shock
accelerated electrons and the emitting electrons cool adiabatically
while the magnetic field in the shell decreases as the shell expands
outwards to larger radii. This naturally accounts for the steep decay
of $\sim t^{-3}$ until $\sim 25\;$days
\citep{gae05,gel05,gra06}. Fig. \ref{fig:dyn_diag} illustrates the
underlying geometry in this model.

As an increasing mass of external medium is swept up, the emission
from the newly shock accelerated electrons within the shocked external
medium rises with time, until eventually (at around $\sim 25\;$days)
it starts to dominate over the rapidly decaying emission from the
merged shocked shell.  When the mass of the swept-up external medium
exceeds that of the merged shell, most of the energy has been
transfered to the shocked external medium and the flow starts to
significantly decelerate, naturally producing a peak in the radio
light curve (at $\sim 33\;$days), followed by a more moderate flux
decay \citep{gel05,gra06}. The fact that the deceleration in the
apparent expansion speed coincides with the peak of the bump in the
radio light curve (see Fig. \ref{fig:sizemod}) nicely supports this
model.

In order to reproduce the observed coasting phase at a constant mildly
relativistic apparent expansion velocity over a factor of $\sim 4-5$
in radius, the bulk of the original outflow (in terms of mass and
energy) could not have been ultra-relativistic, and must have instead
been only mildy relativistic with a velocity very close to that
observed during the coasting phase, i.e. $\sim 0.7c$ at the leading
edge \citep{gra06}. This also implies a large baryonic mass ($\gtrsim
10^{24.5}\;$g) in the outflow which, if spread uniformly over the
outflow, would have obscured the first $\sim 30\;$s of the pulsating
tail of the GF. The fact that such an obscuration did not occur
suggests an anisotropic distribution of baryons in the outflow, where
our line of sight was relatively baryon-poor \citep[and
radiation-rich, in order to see a bright initial spike;][]{gra06}. A
similar requirement arises in order to produce the quasi-thermal
initial spike \citep{NPS05}. This could be manifested, e.g., either if
the baryons are concentrated in a large number of clumps or by some
more ordered global configuration of the outflow (see panels $b$ and
$c$ in Fig. \ref{fig:dyn_diag}).

\subsection{Linear Polarization}

\begin{figure}
%%\epsscale{0.2}
%%\plotone{finf.eps}
%\special{psfile=f1.eps hoffset=-10 voffset=-50 hscale=80.0 vscale=80.0}
\vspace{-1.0cm}
\plotone{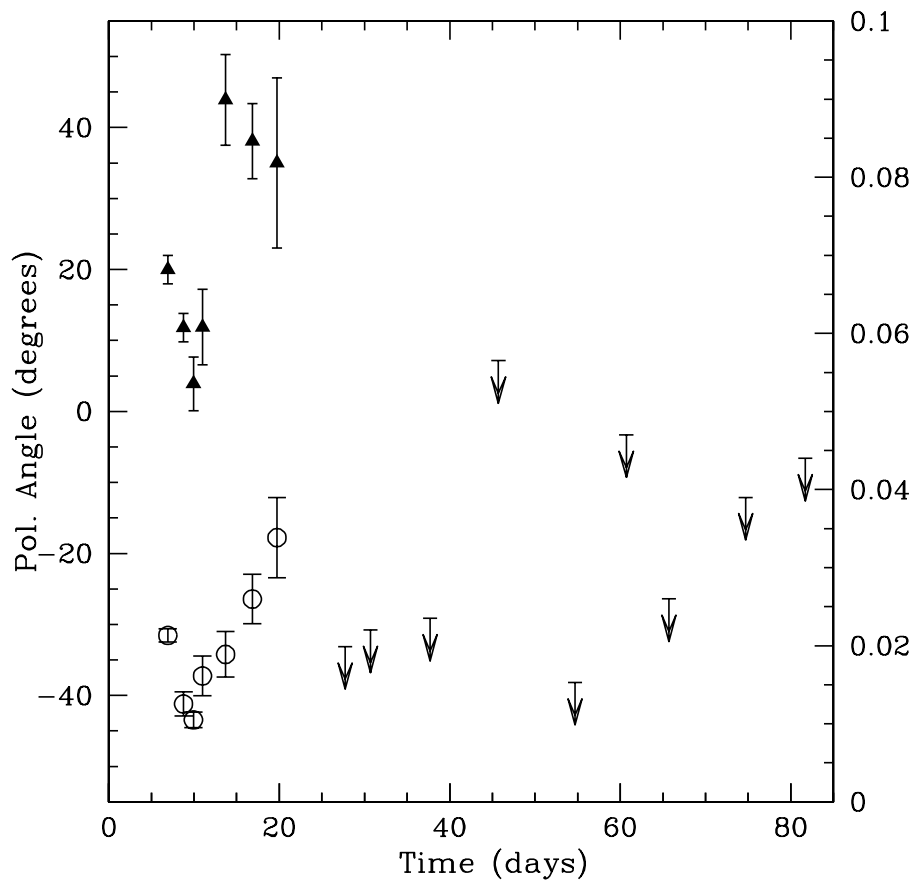}
\vspace{-1.8cm}
\caption{Linear fractional polarization (circles; right y-axis) and
polarization angles (triangles; left y-axis) for the radio afterglow
of the giant flare from \sgr as a function of time at $8.5\;$GHz
\citep[from][]{tay05}.  All polarization angles have been corrected
for the observed RM of $272 \pm 10\;{\rm rad\; m^{-2}}$ \citep{gae05}.
Limits on fractional polarization are drawn at $3\sigma$.}
\label{fig:pol}
\end{figure}

Linear polarization from the radio afterglow was detected during the
first 20 days after the GF at 8.5 GHz \citep{gae05,tay05}. Thereafter
only upper limits on the polarization \citep{tay05} could be set (see
Fig.~\ref{fig:pol}).  The polarization is found to be 2.1\% on day 7
and it decreases to a minimum of 1.1\% on day 10.  At that time the
linear polarization began to increase steadily up to a maximum value
of 3.4\% on day 20 while the polarization angle swung rapidly from
4$^\circ$ to 40$^\circ$.  The polarization falls below our detection
limit of 2\% around the time of the rebrightening in the light curve.
Limits as late as 55 days after the GF are below 2\%
\citep{tay05}. The measured linear polarization and the spectral shape
strongly suggest that synchrotron radiation dominates the radio
emission.

During the first 20 days of high polarization, the emission is
attributed to the shocked ejecta and a shocked external shell
\citep{gae05,gel05,gra06}. If the emission is mostly from the shocked
ejecta, then the degree of polarization of a few percent suggests that
the magnetic field in the ejecta is not dominated by a magnetic field
component ordered on large scales, but is instead tangled on
relatively small scales. A similar conclusion is reached for GRB
outflows, from `radio flare' observations \citep{gt05}. Alternatively,
if the emission is dominated by the shocked external shell
\citep[as suggested by the dynamics;][]{gra06} then the degree of
polarization of a few percent might suggest that the doubly shocked
material in the external shell has a magnetic field that is not
predominantly ordered on large scales.

The degree of polarization decreased around the same time when the
emission started to be dominated by the shocked external medium.  This
suggests a lower degree of polarization in this component,
and in turn that the magnetic field in the shocked external medium is
less ordered than that in the shocked ejecta and/or in the shocked
external shell \citep{tay05}.

The position angle of the linear polarization is roughly perpendicular
to the major axis of the image and to the direction of motion of the
flux centroid. Because of the elongated shape of the emitting region
and due to projection effects \citep{gae05}, such a polarization may
naturally arise for a shock-produced magnetic field. This assumes that
such a magnetic field is tangled predominantly within the plane of the
shock, as expected from a simple linear stability analysis
\citep{ml99}, and manages to survive in the bulk of the shocked fluid
(which is not obvious). Alternatively, such a polarization might be
caused by shearing motion along the sides of the one-sided outflow,
which can stretch the magnetic field in the emitting region along its
direction of motion.

\subsection{Beaming and Energetics}

It is usually argued that, unlike GRBs, the initial $\gamma$-ray spike
of GFs is not significantly beamed. The main argument for this is as
follows. The strong modulation of the tail emission with the
rotational period of the neutron star implies that it is emitted by
material that is confined to the neutron star and co-rotates with it.
This argues against strong beaming of the tail emission (although some
degree of anisotropy is still required in order to produce the
observed pulsations). Furthermore, the pulsating tail of the GFs from
Galactic SGRs (or SGR~0526$-$66 in the LMC) is bright enough to be
detected even without the initial spike. Nevertheless, there is no
observed pulsating tail without a bright initial spike (with an
isotropic equivalent energy output at least comparable to that in the
tail). Such ``spikeless tails'' should, however, be observed if the
initial spike of GFs was strongly beamed into a solid angle
$\Delta\Omega < 4\pi$ and had a negligible (isotropic equivalent)
luminosity outside of this solid angle, for lines of sight outside of
$\Delta\Omega$ (in fact, they should even be more frequent than the
observed GFs which have an initial spike, for a significant beaming
where $\Delta\Omega < 2\pi$).

However, one should keep in mind that in practice such a simple
picture might not be very realistic, and if the luminosity outside of
$\Delta\Omega$ was smaller than inside $\Delta\Omega$ by a large but
finite factor, $f_L \gg 1$, rather than being totally negligible, then
this might explain the lack of ``spikeless tails'', as well as the
difference in the isotropic equivalent luminosity and energy in the
initial spike of the GF from SGR~1806$-$20 compared to that of the two
previous GFs. The peak isotropic equivalent luminosity of initial
spike of the GF from SGR~1806$-$20 was several hundred times larger
than that of the previous two GFs, suggesting that $f_L \sim
10^2-10^3$. The current event rate statistics (one out of three giant
flares observed so far from within $\Delta\Omega$ under this
interpretation) suggest $4\pi/\Delta\Omega \lesssim 100$. Therefore,
there is a significant uncertainty on the degree of beaming of the
initial spike, which implies a similar uncertainty on the true energy
that was radiated during the GF (by a factor of up to $\sim
100$).

There is, however, a somewhat better handle on the kinetic energy of
the outflow from the radio afterglow. The expanding radio nebula
provides a more robust calorimeter for the kinetic energy output of
the GF.  Reproducing the observed synchrotron flux and the size of the
radio nebula around the peak of the bump in the light curve at $\sim
33\;$days provides lower limits on the energy ($E \gtrsim
10^{44.5}\;$erg) and mass ($M \gtrsim 10^{24.5}\;$g) of the outflow,
as well as on the external density ($n \gtrsim 10^{-2.3}\;{\rm
cm^{-3}}$) \citep{gel05,gra06}.  Since the source size and velocity
are measured directly, only the external density $n$ is missing in
order to determine the total mass $M$ and energy $E$ (which both scale
linearly with $n$). The lower limits above correspond to the minimal
energy that produces the observed synchrotron flux when the fraction
of internal energy in the relativistic electrons ($\epsilon_e$) and in
the magnetic field ($\epsilon_B$) in the shocked external medium reach
equipartition values. An upper limit on $n$, $M$, and $E$ may be
obtained by the requirement that the synchrotron self-absorption
frequency is below $\sim 240\;$MHz at $\sim 30\;$days, as implied by
low frequency radio observations \citep[Gelfand et al. in
prep.;][]{cam05b}, and that $\epsilon_B \gtrsim 10^{-3}$ (or that
$\epsilon_e \gtrsim 0.025$): $n\lesssim
0.5(\epsilon_B/10^{-3})^{-0.4}\;{\rm cm^{-3}}$, $M \lesssim
10^{26.5}(\epsilon_B/10^{-3})^{-0.4}\;$g, and $E \lesssim
10^{46.5}(\epsilon_B/10^{-3})^{-0.4}\;$erg (Gelfand et al. in prep.).

\section{Future Work}
\label{sec:future}

Recent A configuration observations with the VLA should provide a good
image of the resolved afterglow one year after the GF.  It will be
interesting to look for signs of limb brightening, or circularization
away from the 2:1 axis ratio seen in the early period of rapid growth.
Owing to the brightness of the afterglow, its slow decay, and
improvements planned for the VLA, this afterglow could potentially be
studied for the next 15 years.  Deep Chandra observations will also
look for the presence of an X-ray nebula.

More detailed modeling of the dynamics of the interaction between the
outflow and its surrounding, including a special relativistic 2D and
3D hydrodynamic calculations are already underway (Ramirez-Ruiz et
al. in prep.), and the effects of magnetic fields are also
considered. Together with better resolution of the radio image with
the VLA in its A configuration, this can provide better constraints on
the properties of the outflow from the GF, and on its immediate
environment.

This event provides a unique opportunity to study the evolution of a
collisionless shock that is initially mildly relativistic, as it
decelerates and becomes increasingly Newtonian. A detailed study of
the radio light curve and spectrum, as well as the evolution of the
source size and morphology, can provide valuable information on the
evolution of the shock microphysical parameters in this interesting
dynamical range around the transition between relativistic and
Newtonian shocks (Gelfand et al. in prep.), bridging the gap between
gamma-ray burst (GRB) afterglows and supernova remnants.

The recent limits from AMANDA-II on the flux of high-energy neutrinos
and photons from the \sgr GF \citep{Achtenberg06} can be used to
constrain the physical properties of the outflow. This could
potentially have interesting implications for the efficiency of
neutrino production and/or the acceleration of UHECRs in the internal
shocks of GRBs.

Ultra-high energy cosmic rays (UHECRs) from the \sgr GF could in
principal arrive at the Earth from its direction years after the
event, and might be detected by AUGER if the deflection of the UHECRs
by Galactic magnetic fields is not too large \citep{asano06}. This can
be tested by AUGER in the years to come.

\acknowledgments 

We thank Bryan Gaensler for helpful suggestions at the outset of this
work, and Ehud Nakar for useful comments on the manuscript.  
GBT thanks Pablo Parkinson at UCSC for inviting him to give the
talk at the SCIPP seminar series on which this review was initially
based. This research was supported by the US Department of Energy
under contract DEAC03-76SF00515 (JG).

\clearpage

\clearpage

\clearpage

\clearpage

\end{document}